# Quasi–1D Nanobelts from the Sustainable Liquid Exfoliation of Terrestrial Minerals for Future Martian-based Electronics

Cencen Wei[1], Abhijit Roy[2,3], Adel K.A. Aljarid[1], Yi Hu[4], S. Mark Roe[5], Dimitrios G. Papageorgiou[4], Raul Arenal[2,3,6] and Conor S. Boland*[1]

[1]School of Mathematical and Physical Sciences, University of Sussex, Brighton, BN1 9QH, U.K.

[2]Instituto de Nanociencia y Materiales de Aragon (INMA), CSIC-Universidad de Zaragoza, 50009 Zaragoza, Spain

[3]Laboratorio de Microscopias Avanzadas (LMA), Universidad de Zaragoza, Calle Mariano Esquillor, 50018 Zaragoza, Spain

[4]School of Engineering and Materials Science, Queen Mary University, London, E1 4NS, U.K.

[5]School of Life Sciences, University of Sussex, Brighton, BN1 9QH, U.K.

[6]ARAID Foundation, 50018 Zaragoza, Spain

*Corresponding Author. Email: c.s.boland@sussex.ac.uk (Conor S. Boland)

**Abstract**

The sky is the limit with regards to the societal impact nanomaterials can have on our lives. However, in this study we show that their potential is out of this world. The planet Mars has an abundant source of calcium sulfate minerals and in our work, we show that these deposits can be the basis of transformative nanomaterials to potentially support future space endeavors. Through a scalable eco-friendly liquid processing technique performed on two common terrestrial gypsum, our simple method presented a cost-efficient procedure to yield the commercially valuable intermediate phase of gypsum, known as bassanite. Through the liquid exfoliation of bassanite powders, suspensions of large aspect ratio anhydrite nanobelts with long-term stability were characterized through scanning electron microscopy and Raman spectroscopy. Transmission electron microscopy showed nanobelts to have a mesocrystal structure, with distinct nanoparticle constituents making up the lattice. Unexpectedly, anhydrite nanobelts had remarkable electronic properties, namely a bandgap that was easily tuned between semiconducting (~2.2 eV) and insulating (~4 eV) behaviors through dimensional control measured via atomic force microscopy. To demonstrate the application potential of our nanobelts; optoelectronic, electrochemical and nanocomposite measurements were made. For the hydrogen evolution reaction and mechanical reinforcement, selenite-based anhydrite nanobelts displayed superlative performances.

**Introduction**

The study of nanomaterials, whether it be one-dimensional (1D) nano-tubes[1] and -wires[2] or two-dimensional (2D) nanosheets,[3] has paved the way in advancing many crucial societal technologies. For a broad range of medical,[4] energy[5] and optoelectronic applications,[6] these nanomaterials have brought about the realization of many *Internet of Things* devices. Recently, 2D nanosheets based on highly abundant, naturally occurring layered minerals were demonstrated to present many surprising properties. Micas and chlorite, from the phyllosilicate mineral family, when exfoliated down to low layer numbers displayed highly tunable electronic properties and unexpected catalytic capabilities.[7] This raises an interesting question of whether other common, abundant minerals could be exfoliated to yield new nanomaterial types that are also optimized for applications. If so, these nanominerals could be a potential source of research inspiration to further galvanize sustainable nanoscience investigations.

One such common mineral of significant commercial interest is gypsum.[8] Globally, more than 150 million tons of gypsum was mined and processed in 2022[9] for a broad range of industries ranging from construction to food supplements.[10] In the US alone, by 2033 the market size for gypsum is expected to be approximately $14 billion.[10] Nonetheless, gypsum in fact can be said to be so common, it has recently been discovered to even inhabit large areas of the Martian surface.[11] On Mars, gypsum is known to exist in large deposits, which are believed to contain most of the primordial water content of the planet.[12] With space exploration a buzz in recent years, humanity has set its sights on Mars as a potential destination. As such, much research has been undergone not only in the way of space travel logistics to our distant neighbor but also on how we might take advantage of the resources it has on offer to sustain human inhabitants. Naturally, one commodity of great interest is gypsum, with recent

NASA funded research exploring modes in which the water content of gypsum may be extracted for human consumption.[13] Through a developed process, gypsum was dehydrated to form water vapor and waste material known as anhydrite. Specifically, anhydrite is the anhydrous phase of gypsum. However, we show here that this waste product could be applied to further sustain and support the future colonization of Mars. Intrinsically, it is known that gypsum minerals are made up of aggregates of amorphous nanobelts.[14] However, beyond understanding how the nanobelts form and coalesce into bulk minerals,[15] little materials science research has been performed on the individual properties of these nanostructures. By demonstrating the exciting potential anhydrite-based nanobelts possess, our work here presents a recourse in which Martian resources could be more fully utilized.

Using a green processing procedure, two forms of raw terrestrial gypsum mineral (selenite and satin spar) through a simple liquid-based dehydration procedure were converted to bassanite. These individual bassanite powders were then liquid phase exfoliated (LPE) in an aqueous medium to yield, regardless of gypsum source, suspensions of large aspect ratio anhydrite nanobelts. Through scanning electron microscopy (SEM), morphological changes occurring during each processing step were examined. We confirmed gypsum's conversion to its anhydrous nanobelt state and the nanobelts long-term stability against ambient reconversion to gypsum through Raman spectroscopy. Nanobelt lattice structure via transmission electron microscopy (TEM) was noted to be mesocrystalline and made up of nanoparticle segments with differing atomic planes. Through cascade centrifugation, suspensions of anhydrite nanobelts were separated into size fractions, revealing the nanobelts to possess bandgaps ($E_g$) as low as ~2.2 eV. Additionally, due to quantum confinement effects in their smallest dimensions, nanobelt $E_g$ rose to values of ~4 eV as width and thickness controllably decreased. In application, anhydrite nanobelt networks displayed light modulated current and catalytic performances for the hydrogen evolution reaction which surpassed many commonly applied nanomaterials. As a filler in a polymer matrix, low loading levels of selenite anhydrous nanobelts displayed properties on par with other nano-reinforcers.

**Results**

Processing and liquid exfoliation of gypsum

Commercially sourced bulk crystal variants of gypsum in the form of selenite and satin spar were procured as raw natural materials. In physical appearance, both minerals greatly differed, namely due to the mechanisms in which they formed. Gypsum minerals are known as *evaporites*,[16] whereby natural bodies of water high in calcium and sulfates have ions precipitate to form elongated bassanite nanoparticles ($CaSO_4 \cdot \frac{1}{2}H_2O$) approximately < 10 nm in length via homogenous nucleation.[14-15, 17] Bassanite nanoparticles then self-assemble to form precursor nanobelt templates that through heterogenous nucleation turn into gypsum nanobelts ($CaSO_4 \cdot 2H_2O$) in the presence of water.[15a, 17] Through variations in growth conditions, tabular, like in the case of selenite, or fibrous, like satin spar, nanobelt aggregates are formed.[15b, 18] From optical images of bulk selenite (Figure 1a) and satin spar (Figure 1b), the difference in growth structure is self-evident. Most noticeably, upon physical examination, was the difference in hue associated with the bulk crystals. For selenite it had a light brown color. While satin spar was milky white and appeared chalky.

As the bulk crystals were raw minerals, a sustainable processing procedure (scheme in Figure 1c, see Methods for more details) was applied to firstly clean them of surface contaminants and to then powder them for liquid exfoliation. The procedure began by grinding the raw bulk minerals in a coffee grinder after which the ground powder was shear mixed in deionized water to remove water soluble contaminants. The resultant solution was then centrifuged, and the sediment taken forward to the next processing step. The next step was the ultra-sonication of a green solvent mixture of sediment in isopropanol (IPA),[19] where the solution was again centrifuged. The sediment was then collected and dried, to yield our clean powder. The clean powder then underwent LPE in an environmentally friendly water/sodium cholate surfactant solution[20] to yield in Figure 1d, high concentration (> 6 mg/mL), stable suspensions of selenite (left) and satin spar (right) nanobelts. What was quite apparent from the appearance of the suspensions was that both liquids were white in hue.

SEM collages revealed morphological changes for selenite (Figure 1e) and satin spar (Figure 1f) during each processing step. As expected, structural differences due to their unique formation mechanisms was self-evident. However, both eventually descended into their basic nanobelt structure upon the completion of processing. For selenite, images of the as received (i) and ground bulk (ii) mineral in Figure 1e presented a layered structure indicative of tabular crystal growth during formation. However, the clean processed powder of selenite (iii) began to reveal the rigid nanobelt aggregates that make up the bulk structure. Through the LPE step (iv), the delamination of the aggregates produced more flexible, individualized species. In contrast, in Figure 1f, satin spar presents a fibrous structure that remained in place for the as received bulk (i), powdered bulk (ii) and cleaned processed powder (iii). The only noticeable difference was the size of the crystals that made up the samples gradually decreased with each step. However, through the LPE step (iv), individual nanobelts were confirmed to have been successfully delaminated from the fibrous crystals.

Material identification

Examining the materials via Raman spectroscopy, we note that for both selenite (Figure 1g) and satin spar (Figure 1h), significant shifting in the vibration $v1$ (a1) mode associated with $SO_4^{2-}$ occurred.[21] Specifically, shifts in this characteristic gypsum mode at ~1008 cm$^{-1}$ appeared across the various stages of mineral processing. Past studies have attributed $v1$ mode shifts with changes in the hydrous state and structural composition of calcium sulfate minerals.[22] In line with these previous reports, we note that for both sets of Raman spectra, the $v1$ mode in the bulk resided at the expected Raman shift of 1008 cm$^{-1}$. However, the mode moved to higher Raman shifts as both materials transitioned to the clean powder state (1014 cm$^{-1}$) and then to the exfoliated nanobelts (1017 cm$^{-1}$). Notably, the $v1$ mode also greatly decreased in intensity with each processing step. We report that these shifts in the $v1$ mode are indicative of the materials transitioning from a dihydrate state in the bulk (*i.e.* gypsum – $CaSO_4 \cdot 2H_2O$) to a hemihydrate state after undergoing cleaning (*i.e.* bassanite – $CaSO_4 \cdot \frac{1}{2}H_2O$) to their anhydrous form when they are exfoliated (*i.e.* anhydrite – $CaSO_4$).[22a] Additionally, the Raman shift position of the transitioning $v1$ mode for each processing step matched precisely previously reported positions for each calcium sulfate phase.

We further confirm the dehydration process by examining the Raman modes of water for each phase. For both bulk selenite and satin spar, water modes at 3404 cm$^{-1}$ and 3492 cm$^{-1}$ were observed.[22a, 23] As anticipated for bassanite minerals, or our clean powder, these modes moved to Raman shifts of 3554 cm$^{-1}$ and 3614 cm$^{-1}$ respectively.[22a] Whereas, for the exfoliated nanobelts or the anhydrous state, modes associated with water are expectedly absent.[22a, 23] We note that even after two months (Figure S1), anhydrite nanobelts do not convert back to bassanite or gypsum through ambient water absorption.[24] This implied that our procedure creates ultra-stable, insoluble anhydrite.[25] X-ray powder diffraction (XRD) spectra in Figure 1i also reflects the compositional changes seen previously via Raman. XRD showed large shifts in peak positions and the appearance of new peaks, confirming a structural change had occurred between the dihydrate bulk mineral and the anhydrous nanobelts.[26] Comparing the anhydrous nanobelts spectra for selenite and satin spar, we note that they are identical with regards to peak positions and shape. This is confirmation that though the crystal structure of the bulk materials differs, the basic building blocks of all calcium sulfate minerals are similarly structured nanobelts. One belief is that bulk calcium sulphate minerals are in fact a mixed phase system, whereby only the external surface is dihydrate gypsum and the internal structure an anhydrous phase composed of nanobelts.[27] This is a plausible explanation for the abundance of anhydrous material generated by LPE. Through X-ray photoelectron spectroscopy (XPS) in Figure 1j, the chemical composition of the LPE nanobelts as expected remained similar to the bulk.[28] We note that peaks associated with Ca, S and O appear in both selenite's and satin spar's anhydrous nanobelt spectra (fully annotated spectra can be found in Figures S2 and S3, respectively). A strong C peak is also noted in both samples, which we attribute to the presence of residual surfactant and ambient organic matter absorbed during crystal formation. Essentially, our mineral processing procedure yields all hydrous forms in which calcium sulfate minerals can exist. Furthermore, it marks a dramatic improvement over energy dense heating regimes[22b] or time intensive ball-milling processes[29] to bring about the dehydration of gypsum to yield its more commercially valuable bassanite or anhydrite forms.

Crystal structure of nanobelts

In Figure 2, we investigated the physical properties of the LPE anhydrite nanobelts through TEM. Here, typical low resolution TEM micrographs of individual anhydrous selenite (Figures 2a and 2b) and satin spar (Figures 2c and 2d) nanobelts are shown. Noticeably in the micrographs, the elongated, belt-like shape of the materials, which align with SEM findings, are evident. However, the surface topography of the nanobelts was more apparent in TEM and appeared uneven, presenting a cross-hatch-like pattern. Nanobelts were composed of a lattice structure with planes perpendicular to the length, indicative of the natural growth mechanism of calcium sulfate structures.[30] Denoted by the red boxes in their respective figures (Figure 4b for selenite and Figure 4d for satin spar) select area electron diffraction (SAED) patterns for the anhydrous nanobelts (insets) were found to be similar to other calcium sulfate phases.[31] High resolution transmission electron microscopy (HRTEM) on the nanobelts revealed both the surfaces of anhydrous selenite (Figure 2e) and satin spar (Figure 2f) to show the expected characteristic polycrystalline structure.[14] Whereby, the overall structure was comprised of bassanite nanoparticles embedded in the lattice. We identified the bassanite nanoparticles through SAED patterns (red boxes), where the expected (204) and (200) planes of bassanite are reported (respective figure insets).[32] Inverse Fast Fourier Transforms (IFFT) of the micrographs in Figures 2e and 2f with respect to the (204) plane (Figure 2g) and the (200) plane (Figure 2h), more clearly revealed the extent of the nanoparticle population. Examining the d-spacing of a select number of colour coded nanoparticles, we confirmed their bassanite nature. We observed the characteristic bassanite lattice spacings of 2.8 Å (Figures 2i and 2j, red lines) and 6 Å (Figures 2k and 2l, red lines),[33] which corresponded to the (204) and (200) planes respectively. Essentially, though Raman shows the overall structure of the nanobelts to be anhydrite, the presence of bassanite nanoparticles in TEM suggests anhydrites may not undergo a full dehydration conversion. As bassanite is believed to be a nucleation point for calcium sulfate phase transitions, the noted sparsity of the bassanite nanoparticle may contribute to long term stability of insoluble anhydrites.[25]

Size selection and statistical analysis of individual nanobelts

Uniquely, with the nanobelts suspended in liquid, it lends towards the application of centrifugation techniques to separate colloids into size fractions of decreasing dimensions.[34] Through this technique, valuable information with regards to size dependent nanobelt attributes can be empirically derived.[35] In Figure 3, a collage of atomic force microscopy (AFM) images shows anhydrous selenite (top) and satin spar (bottom) nanobelt samples that have been drop-casted onto a silicon substrate as a function of relative centrifugal force (RCF). Noted for both sample sets, as RCF value increased from 0.12 g to 0.48 g, the apparent length and height profiles of the nanobelts began to decrease. This is a common occurrence noted previously for other nanobelt systems[36] and nanosheets.[37] In Figure 4, we quantitatively assess the dimensions of the nanobelts for different RCF value samples through statistical AFM. For the nanobelts, we described their shape in terms of their longest dimension (length, $L$), the $L$'s perpendicular bisector (width, $W$) and the profile height of the nanobelt at the $L$'s midpoint (thickness, $t$). AFM histograms of $L$ (Figures 4a and 4b), $W$ (Figures 4c and 4d), and $t$ (Figures 4e and 4f) of anhydrous selenite and satin spar showed that each dimension progressively decreased with RCF. For anhydrous selenite in Figure 4g, mean length ($<L>$) decreased from ~6000 nm for an RCF = 0.12 g to ~1100 nm at RCF = 0.48 g. Similarly, over the same RCF range, mean width ($<W>$) and mean thickness ($<t>$) decreased from ~330 nm to ~140 nm and ~110 nm to ~40 nm, respectively. As a function of $t$ in the same figure, $L$ and $W$ were observed to follow individual power-law scalings with exponents of 2 and 1 respectively. In Figure 4h, a similar behavior in nanobelt dimensions was seen for anhydrous satin spar, with $<L>$ and $<W>$ decreasing from ~4300 nm to ~1300 nm and ~360 nm to ~200 nm, respectively. Furthermore, $<t>$ decreased from ~110 nm to ~40 nm, which we note to be the same value range observed for the selenite-based nanobelts. This implies that during LPE, nanobelts are more prone to scission rather than debundling. This is also reflective of the mesocrystal planes of the nanobelts appearing to being perpendicular to $L$ in TEM micrographs in Figure 2.

However, as a function of $t$, $L$ and $W$ in Figure 4h scaled according to unique power-law scalings with exponents of 1.4 and 0.8 respectively. Plotting anhydrous selenite aspect ratio ($L/t$) as a function of $t$ in Figure 4i, values proportionally scaled with $t$ from ~60 to ~28 as RCF values increased from 0.12 g to 0.48 g. For anhydrous satin spar in Figure 4j, $L/t$ decreased from ~43 to ~25 over a similar RCF value range according to a $t^{0.7}$ scaling. We report that the aspect ratio of the nanobelts were on par with LPE nanosheets (typically < 100)[38] but smaller than carbon nanotubes that were debundled in the liquid phase (typically >250).[39]

Optical and electronic properties of nanobelts

Through UV-Visible spectroscopy (UV-Vis), the light absorption properties of the anhydrous selenite (Figure 5a) and satin spar (Figure 5b) nanobelt suspensions were examined via normalized extinction spectra as a function of RCF value. For both materials, the threshold for absorption shifted to smaller wavelengths as RCF values increased. Through AFM analysis, these spectral shifts can also be interpreted as absorption properties decreasing with nanobelt dimension. As bulk gypsum is reported to be a wide $E_g$ insulator, these findings imply that unexpected electronic transitions were occurring for the exfoliated nanobelts. In Figures 5c and 5d, UV-Vis spectra were converted to Tauc plots to extrapolate information about the optical $E_g$ of each nanobelt species as a function of RCF. Though extinction spectra are comprised of components associated with absorption and scattering effects,[40] Tauc plots from extinction spectra still provide accurate predictions of $E_g$ for large aspect ratio rod-like[41] and disk-like[42] nanomaterials. In Figure 5c, $E_g$ for anhydrous selenite was observed to increase from ~2.24 eV to ~3.93 eV as RCF values rose from 0.12 g to 0.48 g. While for anhydrous satin spar in Figure 5d, $E_g$ increased from ~2.67 eV to ~3.86 eV for the same RCF range. Both minimum $E_g$ values here are far below the expected bulk value of >5 eV.[43] Previous reports on the delamination of other mineral types showed that the initial transition from insulating to semiconducting behavior was due to lattice relaxation.[44] For LPE phyllosilicate mineral nanosheets,[7] this manifested as XPS peak shifts and XRD peak narrowing. However, these findings do not appear to hold true for gypsum, as no consistent data shifts were observed. This is likely due to a complexity of mechanisms associated with transitioning from bulk gypsum to anhydrite nanobelts. However, it has been reported that high levels of residual stress can be found in gypsum and its derivatives due to the natural crystal formation process,[26b, 45] and the presence of contaminates.[46] Essentially, strain relaxation during delamination could plausibly still be the root cause of the initial $E_g$ drop. Nonetheless, strain relaxation does not account for the total occurrence of $E_g$ tunability. When plotting $E_g$ versus $<t>$ and $<W>$ in Figure 5e, values for $E_g$ were seen to scale with an exponent of -2 for both data sets in accordance with quantum confinement.[47] Similarly, phyllosilicate nanosheet $E_g$ was also observed to follow a analogous scaling.[7]

Applications

Previously, semiconducting nanomaterials like transition metal dichalcogenides have shown great promise as optoelectronic devices.[48] With the unexpected semiconducting properties of the nanobelts here, we examined the photoresponse of spray printed networks of anhydrous satin spar on a glass substrate. Due to their large lateral size, the nanobelts making up the printed network were visible via an optical microscope in Figure 6a. For the figure's inset, the long rod-like shape of the nanobelts was self-evident. In Figure 6b, current – voltage (*I-V*) curves associated with the network were investigated under dark and illuminated conditions for a range of monochromatic light sources (650 nm, 532 nm and 380 nm). It was observed that for *I-V* curves under dark conditions or excitation wavelengths with energies below the nanobelt $E_g$ of ~2.6 eV, all curves had similar slopes over a ±5 V range. Furthermore, these curves presented a near linear trend indicative of a resistive element. However, when light with a wavelength of 380 nm (~3.3 eV) illuminated the network, the open circuit voltage ($V_{OC}$) shifted from -1.231 V to 0.276 V. In the forward-bias region, the change in $V_{OC}$ resulted in a crossover voltage of ~2 V between the 380 nm illumination and the other curves. Most noticeably the shape of the 380 nm curve in this region appeared as that of a diode, with a knee point beginning to appear at ~3.3 V and the current showing saturation. Furthermore, the slope in the reverse-bias drastically increased, making anhydrous satin spar nanobelts potentially interesting for photosensitive Zener diodes.

Owing to their exciting electronic properties and rough amorphous surfaces that are fit for reaction nucleation, the potential for hydrogen propagation of anhydrite nanobelts was examined. Electrochemical electrodes were formed by filtering nanobelt suspensions of anhydrous selenite and satin spar on to a membrane, after which they were transferred onto glassy carbon (GC). Polarization curves in Figure 6c for the two nanobelt electrodes presented onset potentials ($V_{onset}$) of 293 ± 21 mV versus RHE and 302 ± 16 mV versus RHE, respectively. While the bare GC electrode had a value of 407 ± 27 mV versus RHE. Through Tafel plots in Figure 6d, the Tafel slope ($S$) values were extrapolated. Anhydrous selenite presented a value of $S = 94 ± 7$ mV/dec, anhydrous satin spar a value of $S = 117 ± 31$ mV/dec and bare GC a much larger slope of $S = 165 ± 19$ mV/dec. Furthermore, derived exchange current density ($J_0$) values for each electrode were 0.0019 ± 0.0002 mA/cm$^2$, 0.0014 ± 0.0006 mA/cm$^2$ and 0.0011 ± 0.0002 mA/cm$^2$ respectively. Comparatively, LPE biotite micene mineral electrodes of similar thickness and density were reported to present a set of metric values of $V_{onset}$ ~ 234 mV, $S$ ~ 95 mV/dec and $J_0$ ~ 0.0035 mA/cm$^2$.[7] In general, we found anhydrite nanobelts to be quite competitive when compared to other nanomaterial types. The nanobelts here greatly surpassed most other nanomaterial catalytic performances, which can be defined as having both low $V_{onset}$ and $T$ values (Figure S5, Table S1). Due to its natural abundance, anhydrous selenite nanobelts in particular prove to be a potential cost-effective replacement for more widely utilized, less sustainably produced active materials.

To take advantage of the large intrinsic stiffnesses associated with single crystal gypsum,[49] mixed-phase anhydrous selenite nanobelt/polyvinyl alcohol (ASN/PVA) nanocomposite materials were made. By creating a range of ASN/PVA nanocomposites with various volume fractions ($V_f$) of filler material, the mechanical reinforcement properties of anhydrous selenite nanobelts were investigated. In Figure 6e, an optical photograph of a typical $V_f \sim 0.2\%$ ASN/PVA sample was shown to be uniform and highly transparent. Through representative stress-strain curves in Figure 6f, variations in the nanocomposite properties as a function of filler loadings was observed. These changes namely manifested themselves in Figure 6g as a linear scaling in Young's modulus ($Y$), from ~34 kPa at $V_f \sim 0\%$ up to 44 kPa at a critical loading of $V_f \sim 0.4\%$. After which, values decreased to ~40 kPa at 0.6 vol% due to filler aggregation.[50] We note that our 1.3 fold increase in modulus at 0.4 vol% was on par with the reinforcement of PVA by graphene oxide (1.5 fold increase at 0.24 vol%),[51] graphene (1.6 fold increase at 0.36 vol%),[52] boron nitride (1.3 fold increase at 0.11 vol%),[53] and molybdenum disulfide (1.1 fold increase)[54] fillers. One other key advantage, unlike the other nanofillers listed above, anhydrous selenite did not affect the opaqueness of the nanocomposites. Using the rule of mixtures and the assumption that the nanobelts are in plane,[55] an estimation of anhydrous selenite ($Y_{AS}$) and composite ($Y_C$) moduli can be made through the expression,

$$Y_C = Y_{AS}V_f + Y_P(1-V_f) \tag{1}$$

Where $Y_p$ is the polymer modulus. Fitting equation 1 in Figure 6g to the low $V_f$ regime, good agreement was found with the data and a $Y_{AS}$ value of 3.6 MPa extrapolated. We note that the $Y_{AS}$ value is far below the potential single crystal theoretical value of ~ 40 GPa,[49a] however this is unsurprising when considering that other mechanical properties remained invariant with $V_f$. In Figures 6h-j, stress at break ($\sigma_B$), strain at break ($\varepsilon_B$) and toughness ($T$) scattered around mean values of 155 kPa, 400% and 350 kJ/m$^3$ respectively. This would suggest that the mechanism for failure in ASN/PVA nanocomposites was the filler/polymer interface.[56] In fact, we find that our evaluation of $Y_{AS}$ was consistent with the shearing of gypsum crystallites.[57] Essentially, stress transfer to the fillers from the polymer matrix results in smaller crystallites being stripped from the nanobelts, causing the interface to fail, rather than the nanobelts fracturing. This is unsurprising when considering the mesocrystal morphology of the nanobelts seen in TEM, where the surfaces are comprised of aggregates rather than a single crystal.[58]

**Discussion**

In conclusion, we demonstrate that naturally occurring bulk terrestrial gypsum can be sustainably processed to produce a wide range of commercially valuable minerals based on its different hydrous phases. Through simple liquid processing techniques, gypsum, bassanite and anhydrite materials can be easily produced. We specifically investigated the properties of anhydrite via liquid exfoliation methods to create water-based suspensions which contained nanobelts that we characterized through a variety of spectral and microscopy techniques. Through these findings, anhydrite nanobelts were discovered to have tunable electronic properties down to semiconducting behaviors that was controlled by the nanobelt's dimensions. With their broad range of superlative properties, nanobelts based on anhydrous selenite and satin spar were demonstrated as effective optoelectronic devices, electrochemical electrodes, and fillers for the mechanical reinforcement of textiles. Our simple methodology and findings lay the foundation to potentially support future exploration and inhabitation of Mars through the usage of its abundant gypsum mineral deposits.

**Methods**

Materials

Bulk gypsum crystals were purchased from Geology Superstore. Selenite (Morocco) and Satin spar (Turkey) were procured as 2-inch x 2-inch samples.

Gypsum processing procedure

A mass of 2 g was removed from the bulk gypsum crystals and powdered using a UUOUU 200 W Bowl Spice Grinder. A 20 mg/mL raw bulk powder/deionised water solution was made up and then shear mixed at room temperature for 1 hour at 5000 rpm. After shearing, the mixture was then centrifuged at 5000 rpm. The supernatant was discarded, and the sediment was redispersed in IPA at 20 mg/mL, firstly by mechanical shaking in hand. The IPA solution was then ultrasonicated (Sonics Vibra-cell VCX130, flathead probe) at 5 °C for 1 hour at 60% amplitude with a 6 seconds on, 2 seconds off configuration. The mixture was then centrifuged at 5000 rpm, with the supernatant again discarded and the bassanite sediment kept. The clean powdered bassanite was then stored in an oven at 60 °C overnight.

## Liquid phase exfoliation

Clean powdered bassanite was added to a sodium cholate solution (Sigma Aldrich BioXtra, ≥99%, 6 mg/mL) at 2 mg/mL and ultrasonicated (Sonics Vibra-cell VCX130, flathead probe) at 5 °C for 5 hours at 60% amplitude with a 6 seconds on, 2 seconds off configuration. The resultant anhydrous nanobelt suspensions were then cascade centrifuged[34a] at RCF values of 0.12 g, 0.21 g, 0.34 g, and 0.48 g.

## Scanning electron microscopy

Samples were drop casted onto a silicon wafer. For the LPE samples a RCF = 0.12 g sample was used for both selenite and satin spar. The wafer was then attached to an SEM stub using a carbon tab and silver paint. All samples were gold coated with a ~5 nm layer. The topography of the samples were examined using a Inspect™ F from FEI Company (Netherlands) in SE2 mode.

## Raman spectroscopy

Suspensions of raw gypsum powder, bassanite powder and anhydrous nanobelts (RCF = 0.12 g) were drop casted onto a glass slide. A Renishaw inVia™ confocal Raman microscope with 0.8 cm$^{-1}$ spectral resolution and 532 nm laser (type: solid state, model: RL53250) was used for measurements. A 2400 mm$^{-1}$ grating in 100× magnification and 5 mW laser power was used. For each sample curve, an average of ten spectra was used.

## X-ray powder diffraction

Powdered bulk and nanobelt suspensions (RCF = 0.12g) were filtered onto nitrocellulose membranes (25 nm pore size). For the nanobelt sample membranes, they were washed with 1 L of deionised water to remove residual surfactant. All membranes after filtering were then left in the oven at 60 °C overnight to dry. A spatula was then used to scrape filter material for the membranes, with the resultant powder collected in a plastic capillary tube. The data was generated on a Rigaku Gemini Ultra using the powder mode in the CrysAlisPro (version 171.42.75) software. The data was collected using Cu radiation (1.5418A) over a range of 125 degrees in 2theta.

X-ray photoelectron spectroscopy

Nanobelt suspensions (RCF = 0.12g) were filtered onto nitrocellulose membranes (25 nm pore size) and washed with 1 L of deionised water to remove residual surfactant. The membranes were then left in the oven at 60 °C overnight to dry. The membranes were measured using a Kratos Axis SUPRA spectrometer utilizing a monochromatic Al K$\alpha$ (1486.6 eV) X-ray source. Survey spectra had an energy step of 1 eV and 160 eV analyser pass energy. The spectra were analysed using CASA XPS software. Shirley baselines were used to subtract the background for quantification purpose. The spectra were calibrated using the B.E. of the C 1s peak at 284.5 eV due to the use of the charge neutralizer during the spectra acquisition.

Transmission electron microscopy

RCF = 0.12 g anhydrous nanobelt suspensions were drop-casted on a 300-mesh carbon coated copper grids and dried. TEM measurements were performed in an image corrected FEI (Thermo Fisher) Titan Cube having a spherical aberration corrector at the objective lens and operating at 300 kV. Electron dose was optimized to reduce any beam induced damage of the samples during the measurements.

Atomic force microscopy

Suspensions were drop casted onto a heated silicon wafer (~70 °C). After which, the wafers were washed with deionised water to remove residual surfactant. A Dimension® icon Bruker positioned in an insulated box over an anti-vibrant stage to minimise environmental noise and building vibrations was used for measurements. For all measurements, a ScanAsyst Air tip probe with a spring constant of 0.4 N m$^{-1}$, and a tip−sample contact force of 5.0 nN was used. To obtain a good statistical average for length, width and thickness; 200 measurements for each dimension was performed through line profile analysis of individual nanobelts.

UV-Vis spectroscopy

Measurements were performed using a Shimadzu UV-3600 Plus spectrophotometer from 200−800 nm using a quartz cuvette (path length, 1 cm). Sample curves were an average of five spectra.

Printed anhydrous satin spar nanobelts

A KMOON airbrush stylist (nozzle diameter = 0.2 mm, fluid cup capacity = 9 cc) was used at a pressure of 3.6 bar to deposit a 1.2 mg/mL suspension of anhydrous satin spar nanobelts onto a glass substrate. Optical images of the network were taken using an Olympus BX53M with 4K digital CCD camera. The thickness of the sample was measured using a Bruker DektakXT profilometry. The uniform 25 mm x 25 mm nanobelt network had silver contacts painted

(length 25 mm) onto the central surface of the network at an electrode distance of 3 mm. A Keithley 2614B voltage source using the voltage range of -5 to 5 V. For all measurements, the sample was covered by a blackened tarp. Lights sources (5 mW laser pen) were positioned ~10 cm away from the sprayed network surface. Standard error for all source wavelengths was ±10 nm.

Electrochemical analysis:

Suspensions (RCF = 0.12 g) of known concentration found via vacuum filtration, were filtered onto nitrocellulose membranes (25 nm pore size) and washed with 1 L of deionised water to remove residual surfactant. Nanobelt networks (density = 0.2 mg cm$^{-2}$ ± 0.013, thickness = 410 nm ± 150 nm) on the membrane were then cut into pieces and transferred onto glassy carbon rods (3 mm diameter, BASi) by placing the membrane (network side down) on the substrate. The membrane was then wetted with isopropanol and pressure applied. Acetone vapour and acetone baths were used to then dissolve the membrane, leaving the bare network on the substrate. Linear sweep voltammetry and electrochemical impedance spectroscopy were performed using a Gamry Reference 3000 potentiostat in a three-electrode configuration. The glassy carbon electrode was used as the working electrode, while a platinum wire and Ag/AgCl (3 M KCl) was used as the counter and the reference electrode, respectively. Linear sweep voltammetry experiments were performed with a scan rate of 5 mV s$^{-1}$ from 0 V to −1.5 V (*vs.* RHE) in 0.5 M $H_2SO_4$ to investigate the hydrogen evolution performance. The measured potential was converted to the RHE scale by adding +0.2 V, measured with respect to a Gaskatel Hydroflex $H_2$ reference electrode.

Anhydrous selenite nanobelt nanocomposites

A 20 mg/mL stock solution of PVA (Sigma-Aldrich, product code: 102415238, CAS-No: 9002-89-5, average $M_w$ between 30,000 and 70,000) was created by mixing PVA powder in deionized water at 180°C at 700 rpm for 1 hour. To make nanocomposites with different mass fractions of filler, various volumes of anhydrous selenite (6 mg/mL) were mixed into a volume of PVA solution, with all samples having a constant total volume of 20 mL. The mixed phase solutions were then mixed at 180°C at 700 rpm for 1 hour. After which, the solution was left to cool ambiently for a few minutes and then poured into a petri dish. The petri dish was then placed in a vacuum oven at 60°C under a vacuum of 900 mbar. Nanocomposite filler volume fraction was calculated from the mass fraction using the densities of the polymer (1190 kg/m$^3$) and the filler (2960 kg/m$^3$). Nanocomposites were cut into 47 mm x 5 mm segments and analyzed using a Stable Micro Systems TA-TXplus at a test speed of 0.6 mm/sec. Thickness of each test segment was measure using a screw gauge.

**Competing Interests**

Authors declare that they have no competing interests.

**Supporting Information**

Supporting Information is available online and includes supplementary text, Figures S1 to S5, Table S1, References 1 – 9.

**Acknowledgements**

C.W. and C.S.B. acknowledge funding from University of Sussex Strategic Development Fund. A.K.A.A. acknowledges funding through the Saudi Arabian Cultural Bureau. Y.H. and D.G.P. acknowledge funding support from "Graphene Core 3" GA: 881603 implemented under the EU-Horizon 2020 Research & Innovation Actions (RIA) and supported by EC-financed parts of the Graphene Flagship. D.G.P. also acknowledges support from The Royal Society through RGS\R2\212410. A.R. and R.A. acknowledges funding from the Spanish MCIN (project grant PID2019-104739GB-100/AEI/10.13039/501100011033), from the Government of Aragon (project DGA E13-20R) and from the European Union H2020 program "ESTEEM3" (823717).

**Author Contributions**

C.S.B. conceived and designed experiments. C.W. created all samples for the study. C.W., A.J., A.K.A.A., Y.H., S.M.R. performed experiments. C.S.B. analysed and modelled data. D.G.P., S.M.R., R.A., C.S.B. contributed materials, tools, and supervision. C.S.B wrote and revised the paper.

**Dedications**

For O.R.B. – You always make each day a special day, by just being you.

# Figures

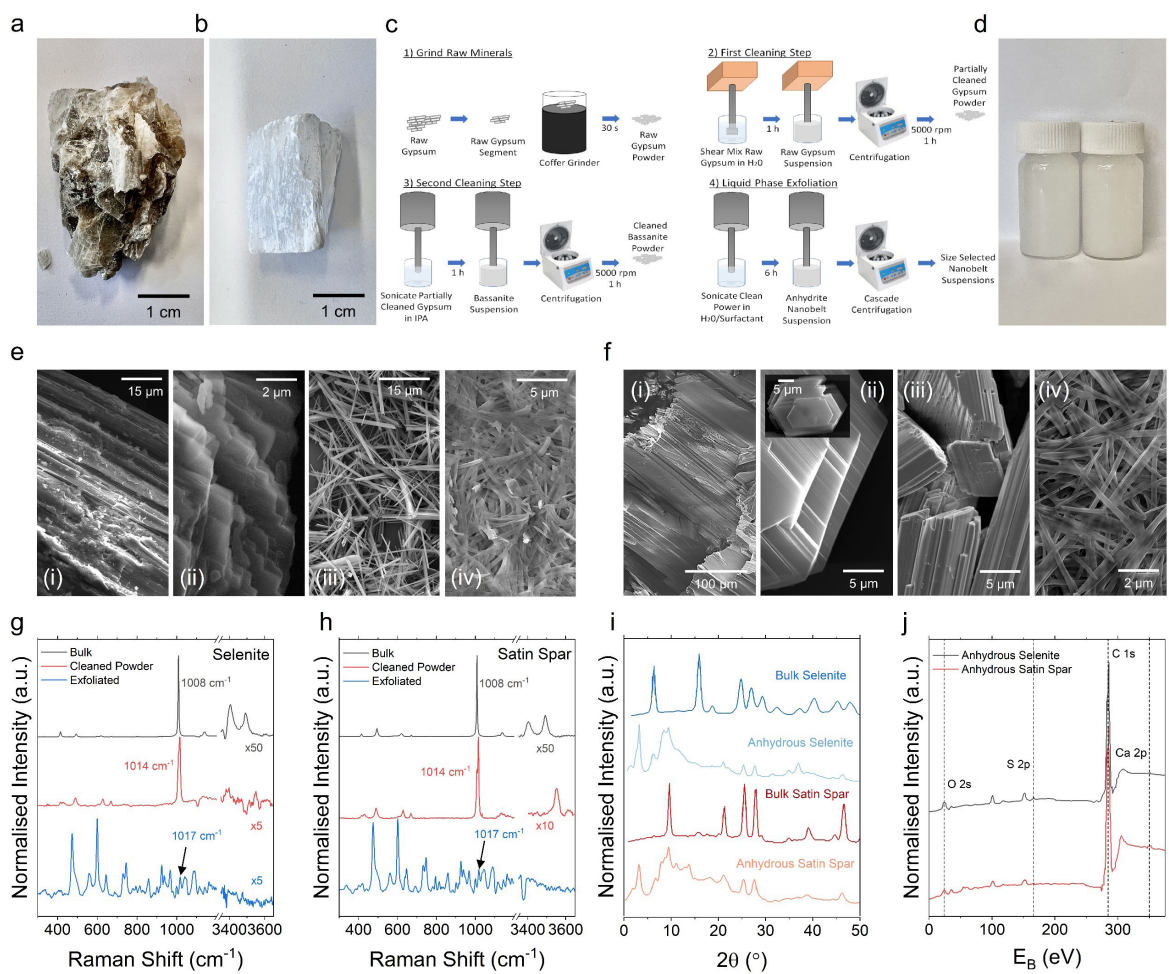

**Figure 1: Processing and Basic Characterisation of Minerals.** Optical photograph of the raw, bulk (**a**) selenite and (**b**) satin spar minerals. (**c**) Graphical scheme depicting the procedural steps applied to process the raw minerals into a powder, which underwent a series of cleaning steps to yield bassanite powder that was liquid exfoliated to create anhydrous nanobelts. (**d**) Optical photograph of an (left) anhydrous selenite and (right) anhydrous satin spar suspension. (**e**,**f**) Scanning electron micrograph collage showing the various morphological stages of (**e**) selenite and (**f**) satin spar mineral during processing and exfoliation. Left to right: (i) as received bulk material, (ii) raw bulk powder, (iii) cleaned powder/bassanite powder, and a (iv) liquid exfoliated anhydrous nanobelt network. (**g**,**h**) Raman spectra of (**g**) selenite and (**h**) satin spar samples during their various stages of crystal structure. (**i**) X-ray powder diffraction spectra comparing the crystal structure change between the bulk powder and the nanobelts. (**j**) X-ray photoelectron spectroscopy confirming the elemental composition of the nanobelt materials. Dashed lines denote peaks related specifically to oxygen, silicon, carbon and calcium.

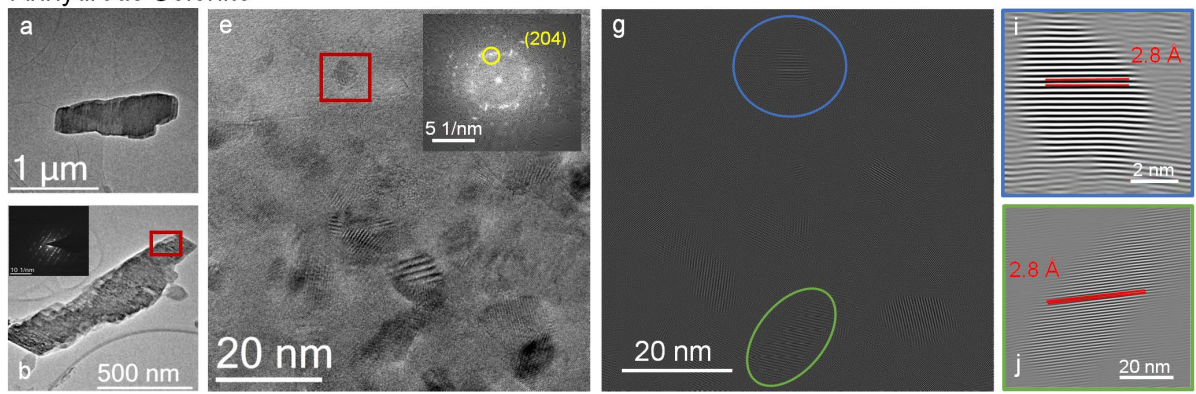
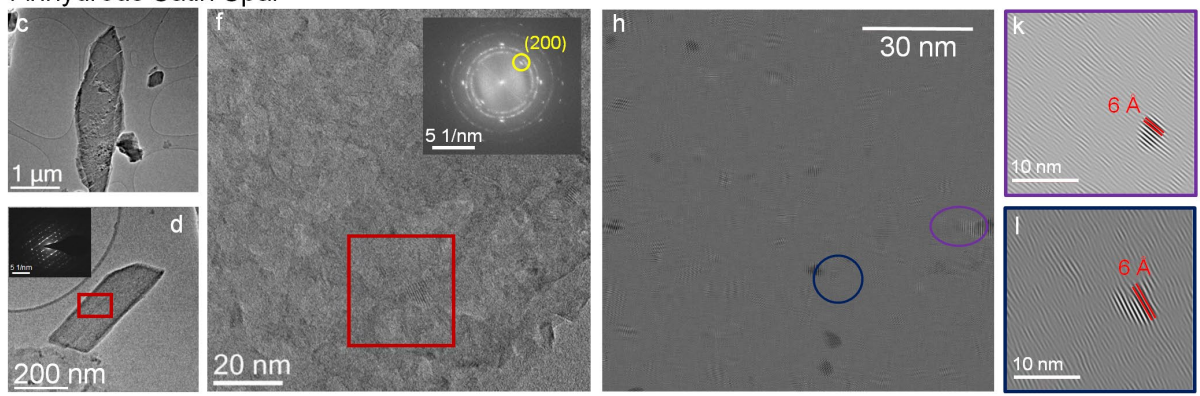

**Figure 2: Nanobelt mesocrystal structure.** (**a-d**) Low resolution transmission electron micrographs of anhydrous (**a,b**) selenite and (**c,d**) satin spar nanobelts. Red boxes in figures **b** and **d** denote the regions in which the respective inset select area electron diffraction patterns were taken. (**e,f**) High resolution transmission electron micrographs showing the lattice composition of the anhydrous selenite and satin spar nanobelts, respectively. Red box in figure **e,f** represents the region in which the select area electron diffraction patterns of bassanite nanoparticles were taken, presenting typical Miller (*hkl*) indices of (204) and (200) respectively. (**g,h**) Inverse Fast Fourier Transform of figures **e** and **f** with respect to the (204) and (200) planes of the bassanite nanoparticles, respectively. A select number of nanoparticles are highlighted by colour coded rings. (**i-l**) The characteristic d-spacing (red lines) for the (204) and (200) planes of bassanite were found for the respective colour highlighted nanoparticles.

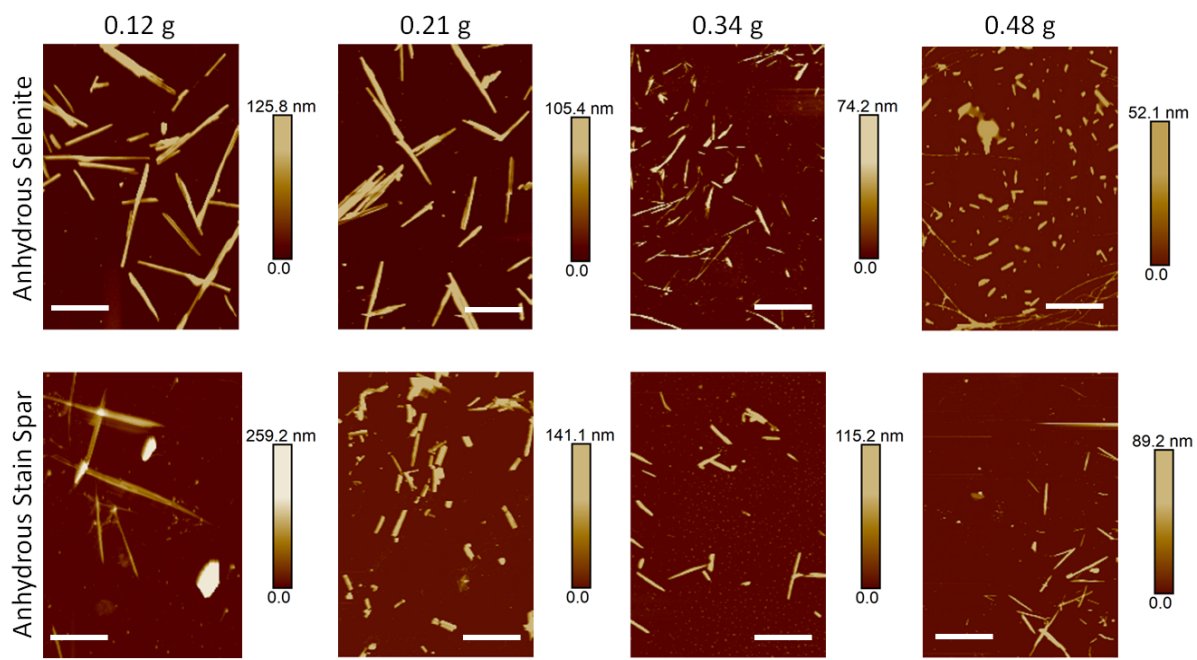

**Figure 3: Size selected nanobelts.** Representative atomic force microscopy images of drop casted nanobelt suspensions on silicon wafers as a function of relative centrifugal force (RCF = 0.12, 0.21, 0.34, 0.48 g). Both anhydrous (top row) selenite and (bottom row) satin spar presented suspensions that had nanobelt dimensions that decreased as a function of increasing RCF. Scale bars equal to 6000 nm.

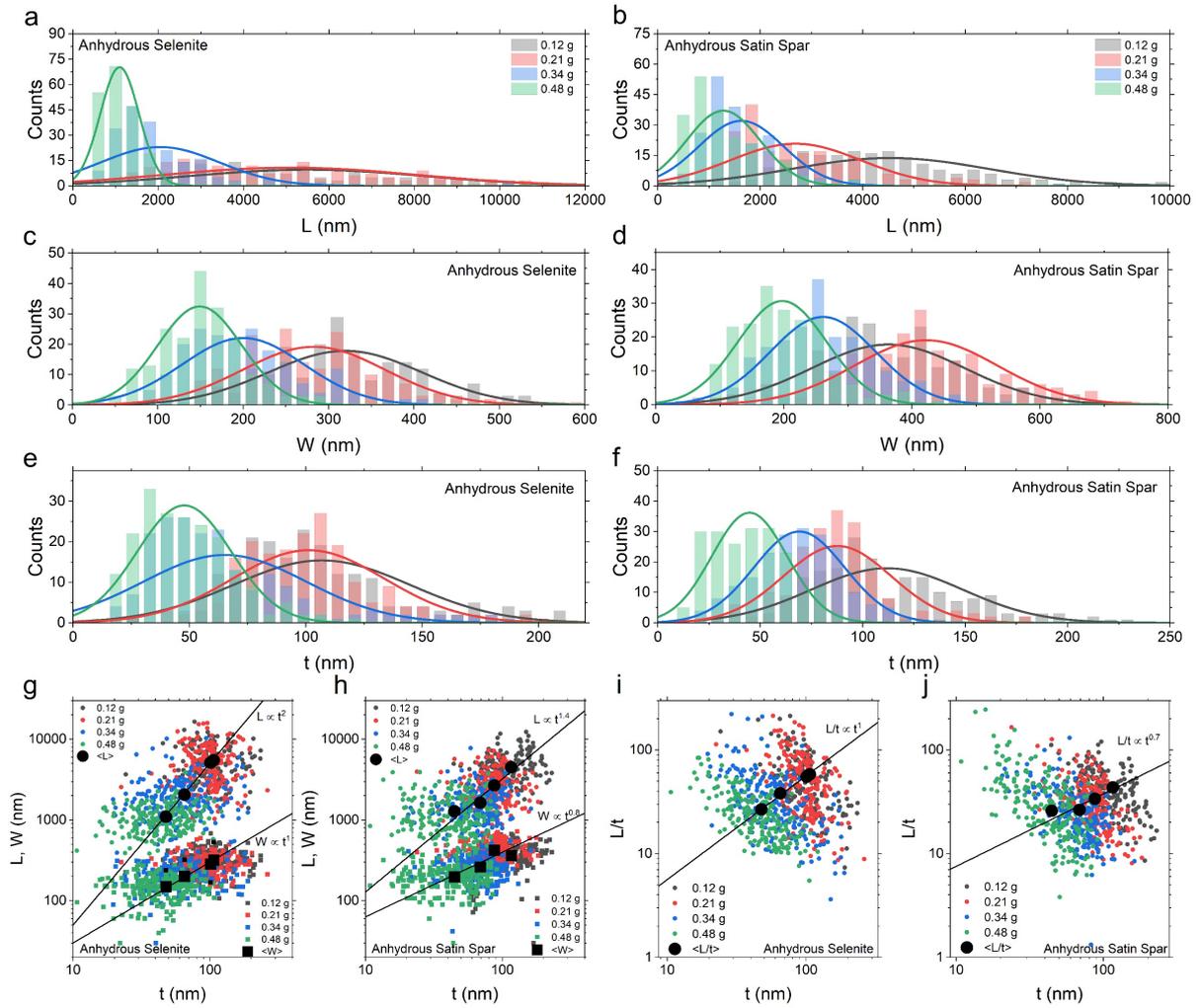

**Figure 4: AFM analysis of size selected nanobelt suspensions.** (**a-f**) Histograms showing statistical atomic force microscopy (AFM) analysis associated with the (**a,b**) length (*L*), (**c,d**) width (*W*) and, (**e,f**) thickness (*t*) of anhydrite nanobelts as a function of relative centrifugal force (RCF). (**g,h**) Length and width of (**g**) selenite and (**h**) satin spar anhydrite nanobelts for a range of RCF values as a function of thickness followed individual power-law like scalings (solid lines). (**i,j**) Aspect ratio (*L/t*) for both (**i**) anhydrous selenite and (**j**) satin spar nanobelts also scaled according to unique power-laws (solid lines). With selenite *L/t* scaling proportionally with nanobelt thickness (*i.e.* exponent of 1) and satin spar *L/t* scaled with an exponent of 0.7 as a function of *t*.

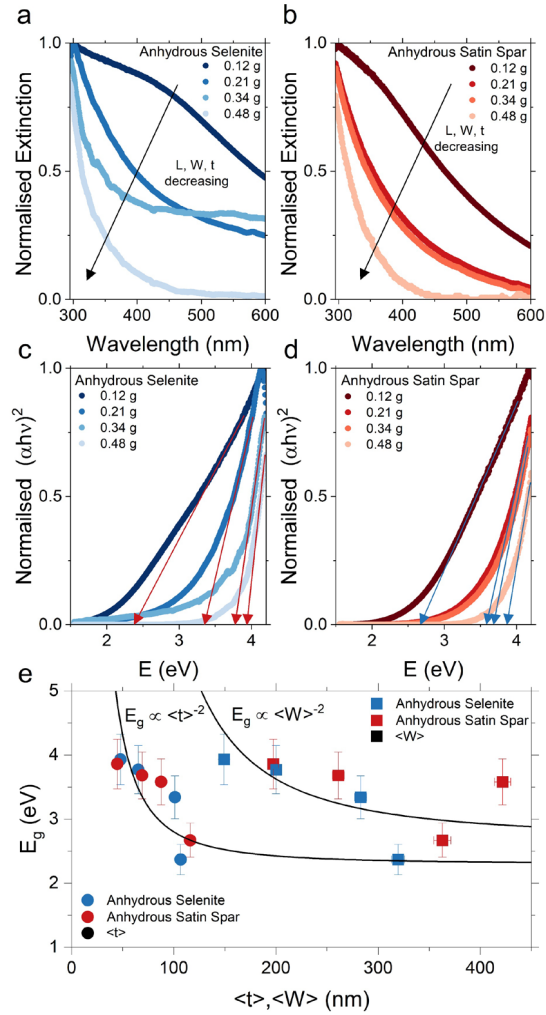

**Figure 5: Size dependent optical and electrical properties.** (**a**,**b**) Normalised UV-Vis extinction spectra of anhydrous selenite and satin spar, respectively, as a function of relative centrifugal force (RCF). As a resultant, spectral lines are noted to scale with decreasing length (*L*), width (*W*), and thickness (*t*) associated with the size selected nanobelt suspensions previously measure via AFM. (**c**,**d**) Tauc plots derived from the UV-Vis data in figure **a** and **b** allow for the calculation of the optical bandgap ($E_g$) for (**c**) anhydrous selenite and (**d**) anhydrous satin spar. The values of $E_g$ for both materials were seen to incrementally increase with RCF value, and thus decreasing nanobelt dimensions. (**e**) Specifically, $E_g$ scaled with both anhydrite nanobelt mean thickness (<*t*>) and mean width (<*W*>) according to quantum confinement effects, defined by <*t*>$^{-2}$ and <*W*>$^{-2}$ scalings (solid lines).

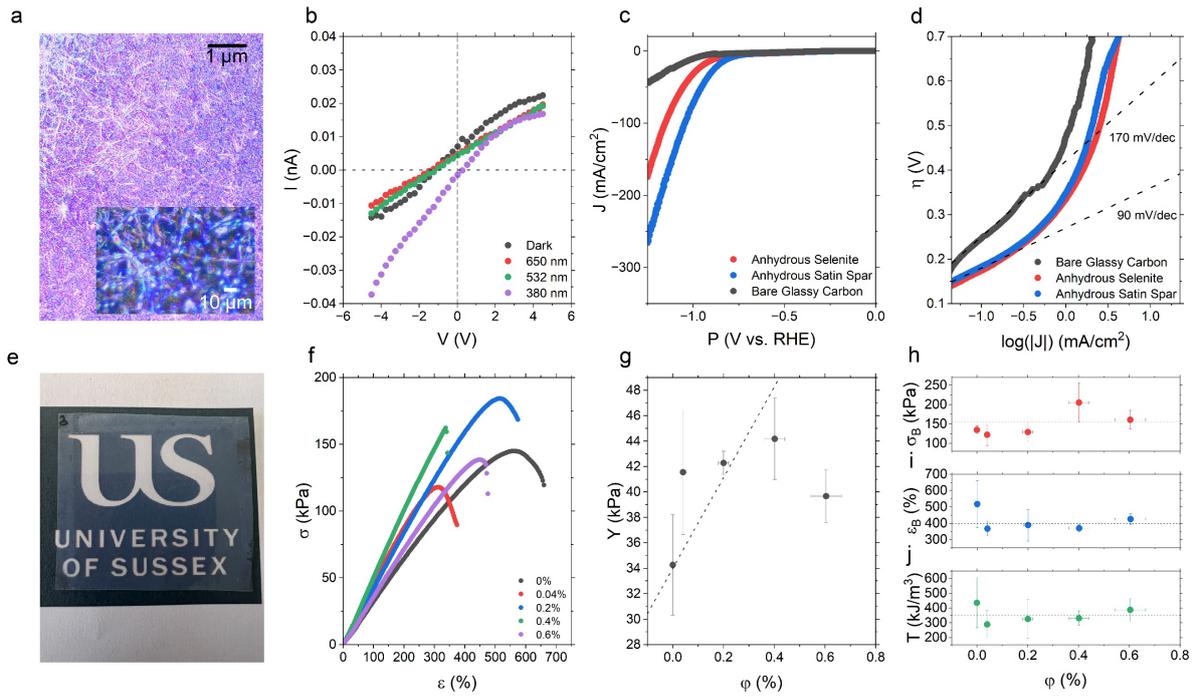

**Figure 6: Applications of nanomineral nanobelts.** (**a**) Optical micrographic image of a 2 μm thick anhydrous satin spar nanobelt network spray printed onto a glass substrate. Inset is a zoomed in look at the network. (**b**) Current verse voltage curves of the anhydrous satin spar nanobelt network from figure **a** in the dark and as a function of various illumination wavelengths. (**c**) Polarisation curves for anhydrite nanobelts and the bare glass carbon electrode, with their corresponding (**d**) Tafel plots. Dashed lines in figure **d** denote the Tafel slope. (**e**) Optical image of 0.2 vol% anhydrous satin spar/PVA nanocomposite on top of the University of Sussex logo. (**f**) Representative stress-strain curves of PVA nanocomposites as a function of anhydrous satin spar loading level. (**g-j**) Extrapolated mechanical metrics Young's modulus ($Y$), stress at break ($\sigma_B$), strain at break ($\varepsilon_B$) and toughness ($T$)) as a function of loading level. Dashed line in figure **g** is a fit of equation 1 and dashed lines in figures **h-j** are mean values of 155 kPa, 400% and 350 kJ/m$^3$ respectively.